\documentclass[12pt]{article}
\usepackage{amsmath,amssymb,graphicx}

\newcommand{\be}{\begin{eqnarray}}
\newcommand{\ee}{\end{eqnarray}}
\newcommand{\nn}{\nonumber}
\newcommand{\ti}[1]{{\tilde{#1}}}

\def\a{\alpha}
\def\o{\omega}
\def\s{\sigma}
\def\cV{{\cal{V}}}
\def\cH{{\cal{H}}}
\def\cP{{\cal{P}}}
\def\cG{{\cal{G}}}
\def\cW{{\cal{W}}}
\def\cQ{{\cal{Q}}}
\def\cL{{\cal{L}}}
\def\cJ{{\cal{J}}}
\def\pt{\partial_t}
\def\vp{\varphi}
\def\L{\Lambda}
\def\E{E_{10}}
\def\KE{E_{10}/K(E_{10})}
\def\lae{{\mathfrak{e}}_{10}}
\def\LRA{\Longleftrightarrow}
\def\cx{{\mathbb{C}}}

\begin{document}

{\flushright{AEI-2005-179}\\[10mm]}

\begin{center}{\LARGE \bf \sc $E_{10}$ Cosmology}\\[15mm]
{\bf Axel Kleinschmidt} and {\bf Hermann Nicolai}
\\[4mm]
Max--Planck--Institute for Gravitational
Physics, Albert--Einstein--Institute, Am M\"uhlenberg 1, 14476
Potsdam, Germany\\[1mm]
Emails: {\tt axel.kleinschmidt,hermann.nicolai@aei.mpg.de}\\[8mm]

\begin{tabular}{p{12cm}}
\hspace{2mm}{\bf Abstract:} We construct simple exact solutions to the
$\KE$ coset model by exploiting its integrability. Using the
known correspondences with the bosonic sectors of maximal
supergravity theories, these exact solutions translate into exact
cosmological solutions. In this way, we are able to recover some
recently discovered solutions of M-theory exhibiting phases of
accelerated expansion, or, equivalently, S-brane solutions, and
thereby accommodate such solutions within the $\KE$ model. We also
discuss the situation regarding solutions with non-vanishing
(constant) curvature of the internal manifold.
\end{tabular}

\end{center}

\begin{section}{Introduction}

Cosmological solutions of Einstein gravity coupled to (axionic
and/or dilatonic) scalar and $p$-form matter have been studied for
a long time. In particular, those systems which arise as the
bosonic sectors of (maximal) supergravities have attracted
attention due to their appearance as low energy effective actions
in string and M-theory (see
\cite{DeHaHeSp85,BeFo94,LuMuPoXu97,IvMe20,CoCo02,ToWo03,Ohta:2003pu,Wo03,Fre,Collinucci:2004iw,Oh04,DeIvMe05,Fre:2005bs}
and references therein). As was shown in
\cite{ToWo03,Ohta:2003pu,Wo03,Oh04} some  of these solutions
exhibit a phase of accelerated expansion. Although these solutions
fail to generate an inflationary period on the scale required by
current observational data \cite{Wo03,Ohta:2003pu,Oh04}, a most
intriguing feature of these results is clearly the fact that M
theory ({\it alias} $D=11$ supergravity \cite{CJS}) admits
cosmological solutions with interesting profiles for the time
evolution of the cosmic scale factor, with intermittent periods of
accelerated and decelerated expansion.

More recently, it has been appreciated that the dynamics of the (spatial)
scale factors and dilaton fields in the vicinity of a space-like
singularity can be effectively described in terms of a cosmological
billiard (for a review and references to earlier work, see \cite{DaHeNi03}),
which in many cases of interest (and for all examples in M- or superstring
theory \cite{DaHe00,DaHeNi03}) takes place in the Weyl chamber of some
indefinite
Kac--Moody algebra. Motivated by this unsuspected link between cosmological
solutions and the theory of (indefinite) Kac--Moody algebras, a more general
framework was developed in \cite{DaHeNi02}, which relates (a certain
truncation of) the bosonic equations of motion of the Einstein-matter
system to a null geodesic motion on the infinite-dimensional coset
space, which is defined as the (formal) quotient of the relevant Kac--Moody
group by its maximal compact subgroup. The example which has been studied
in most detail is the coset space $\KE$, where $E_{10}$ is a maximally
extended (rank 10) hyperbolic Kac--Moody group $\E$, and $K(\E)$ its maximal
compact subgroup. The equations of motion of the corresponding $\KE$
$\sigma$-model can be expanded in terms of `levels' defined w.r.t. certain
distinguished finite-dimensional subalgebras of $\E$. When this expansion
is truncated to low levels, the resulting (truncated) equations of motion
can be shown to correspond to a truncated version of the bosonic equations
of motion of $D=11$ supergravity \cite{DaHeNi02,DaNi04}, massive IIA
\cite{KlNi04a} and type IIB supergravity \cite{KlNi04b}, respectively,
where the level is defined w.r.t. the $A_9$, $D_9$ and $A_8\times A_1$
subgroups of $\E$, respectively.\footnote{In the framework of the 
$E_{11}$ proposal of \cite{We01}, the common origin of these theories
was already discussed in \cite{West:2004st}.} The truncation of the 
supergravity field equations here corresponds to a dimensional reduction 
to one (time) dimension, where, however, first order spatial gradients are 
kept. The parameter of the coset space geodesic in this context is interpreted
as the time coordinate of the Einstein-matter system and all fields depend
on it. According to the main conjecture in \cite{DaHeNi02}, some of the
higher levels will then correspond to higher and higher order spatial
gradients, in such a way that the dependence on the spatial coordinates
(which seems to be lost in a na\"ive reduction to one dimension) gets
`spread over' the infinite-dimensional Kac--Moody Lie algebra. However,
the vast majority of higher level representations are expected to
correspond to new, M-theoretic, degrees of freedom which have no counterpart
in the effective low energy supergravity theory. For some ideas on the
r\^ole of imaginary root generators we refer to
\cite{BrGaHe04,DaNi05}.

The aim of this letter is to show that the hyperbolic $\KE$ $\sigma$-model
of \cite{DaHeNi02} naturally includes accelerated cosmological
solutions of the type studied in \cite{ToWo03,Ohta:2003pu,Wo03,Oh04}.
A crucial new ingredient in our analysis is the incorporation of
{\em Borcherds}-type subalgebras of $E_{10}$, which are characterized
by the presence of {\em imaginary simple} roots.\footnote{There are many
 Borcherds subalgebras inside $\lae$, with an arbitrarily large number
 of imaginary simple roots, see e.g. \cite{FN}. The complexifications
 of the rank-one Borcherds algebras considered below (for $\a^2\ne 0$)
 are still isomorphic to $\mathfrak{sl}_2(\cx)$, but with a compact
 (negative norm) Cartan generator $h$ over the real numbers.}
Our strategy will be to exploit the integrability of the $\KE$ model.
For clarity we restrict attention to the simplest cases as it will
turn out that these already produce the solution given in
\cite{Wo03} and outline at the end how to obtain more general solutions
of cosmological type, as this generalisation is straightforward.

We note that, in the context of Kac--Moody theoretic approaches to
supergravity, various types of solutions had already appeared in 
the literature before \cite{West:2004st,Cook:2004er,Englert:2003py,EnHo04b}. 
Although, like previous solutions, our solutions are also obtained 
from a non-linear realization of a Kac--Moody group, they are of 
{\em cosmological type}, i.e. time-dependent, and thus differ from the 
static BPS solutions found in \cite{West:2004st,Cook:2004er} (and
cannot be obtained from those by a formal Wick rotation).
Another main difference between the approach taken here and the one in
\cite{We01,West:2004st,Cook:2004er} is that we obtain exact solutions 
to an abstract {\em one-dimensional} coset model, and then use the known 
(partial) correspondences with different supergravity models to interpret 
these abstract solutions in cosmological terms. On the other hand, 
the exact $E_{11}/K(E_{11})$ geodesic coset model solutions 
of~\cite{Englert:2003py,EnHo04b}, when interpreted in a cosmological 
context as in \cite{Englert:2004ph}, are similar to our approach.

This letter is structured as follows. In section \ref{esec} we
define the $\KE$ coset model and demonstrate how to obtain exact
solutions to this integrable system in section \ref{solsec}.
Using the established Kac--Moody/supergravity dictionary,
we translate one particular such
solution to $D=11$ supergravity in section \ref{accsec}. At the
present stage there are difficulties with accommodating related
gravity solutions with highly curved spaces as will be discussed
in section \ref{nonflat}. Our conclusions are presented
in section \ref{concsec}.

\end{section}

\begin{section}{Cosmological Solutions from $\E$}

\begin{subsection}{$E_{10}$ model}
\label{esec}

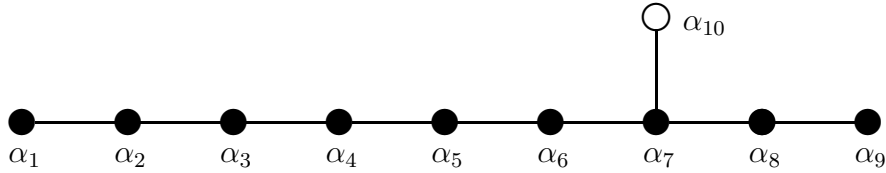
\begin{figure}[t!]
\begin{center}
\scalebox{1}{
\begin{picture}(340,60)
\put(5,-5){$\alpha_1$}
\put(45,-5){$\alpha_2$}
\put(85,-5){$\alpha_3$}
\put(125,-5){$\alpha_4$}
\put(165,-5){$\alpha_5$}
\put(205,-5){$\alpha_6$}
\put(245,-5){$\alpha_7$}
\put(285,-5){$\alpha_8$}
\put(325,-5){$\alpha_9$}
\put(260,45){$\alpha_{10}$}
\thicklines
\multiput(10,10)(40,0){9}{\circle*{10}}
\multiput(15,10)(40,0){8}{\line(1,0){30}}
\put(250,50){\circle{10}} \put(250,15){\line(0,1){30}}
\put(290,10){\circle*{10}}
\end{picture}}
\caption{\label{e10dynk}\sl Dynkin diagram of $\lae$. The solid
nodes correspond to the regular $\mathfrak{sl}(10)$ subalgebra
which can be extended to $\mathfrak{gl}(10)$.}
\end{center}
\end{figure}

The hyperbolic Kac--Moody algebra $\lae$ is defined by the Dynkin
diagram of fig.~\ref{e10dynk} \cite{Ka90}. An element of the
corresponding group coset $\KE$ is  given by an Iwasawa-like
parametrisation
\be
\label{cosel}
\cV (t)= \exp\left(\sum_{\a\in\Delta_+} A_\a(t) E_\a\right)
\exp\left(\sum_{i=1}^{10} \phi_i(t) K^i{}_i\right)
\ee
Here, $E_\a$ are the positive root (upper triangular) generators of $\lae$
(as is customary, we denote by $\Delta_\pm$ the set of positive and negative
roots, respectively). Since we will be interested in the $\mathfrak{gl}(10)$
decomposition of $\lae$, we have already adopted a basis of ten
diagonal $GL(10)$ generators $K^i{}_i$ as a basis of the Cartan
subalgebra of $\lae$. (For details on the decomposition see
\cite{DaHeNi02,NiFi03}.)

The $\KE$ coset model is constructed from (\ref{cosel}) in the standard
fashion. Associated with $\cV$ is the velocity
\be\label{cartform}
\cV^{-1}\pt\cV = \cP + \cQ,
\ee
where $\cQ$ corresponds to the unbroken local $K(E_{10})$ invariance of
the model, and is therefore fixed by the Chevalley involution $\o$
defining the invariant subalgebra ${\rm Lie}(K(E_{10}))$. The remaining
components $\cP$ are along the coset directions in $E_{10}$. They contain
the physical fields of the model. The time reparametrisation invariant
Lagrangian is given by
\be\label{sigmod}
\cL = \frac1{2n}\langle \cP | \cP \rangle,
\ee
with the standard invariant bilinear
form $\langle\cdot|\cdot\rangle$ \cite{Ka90}. The equations of
motion implied by this Lagrangian are
\be\label{eomsig}
n\pt(n^{-1} \cP) = [\cP, \cQ],\quad\quad\quad \cH\equiv\langle \cP
| \cP\rangle = 0.
\ee
the latter being the Hamiltonian constraint associated with the
invariance under time reparametrisations. The system is formally
integrable since the $\lae$-valued Noether current
\be
\cJ = n^{-1}\cV \cP \cV^{-1}
\ee
contains infinitely many conserved charges. The formal solution of
(\ref{eomsig}) is
\be\label{gensol}
\cV = \exp(\nu(t)\cJ)\,\cV_0\, k(t)  \qquad \mbox{with} \quad
\nu(t) := \int_{t_0}^t n(t') dt'
\ee
for any choices of charges $\cJ$ and initial configuration
$\cV_0$; the compensating $K(E_{10})$ rotation $k(t)$ is determined
so as to bring $\cV$ into the triangular gauge (\ref{cosel}), and depends
on both $\cJ$ and $\cV_0$.

\end{subsection}

\begin{subsection}{Construction of $\KE$ Solutions}
\label{solsec}

Even though the general solution to the $\KE$ model is known,
the usefulness of (\ref{gensol}) is limited by (at
least) two facts. First, the structure of $E_{10}$ is not known in
closed form which complicates the evaluation of (\ref{gensol}), in
particular the determination of $k(t)$ --- even disregarding
ambiguities in defining the Kac--Moody groups. Secondly, the
relation of the $\KE$ model to $D=11$ supergravity is presently only
understood for a truncation of the system (\ref{eomsig}) to low levels
\cite{DaHeNi02,DaNi04}. Hence, an interpretation of the general
solution (\ref{gensol}) in terms of supergravity or M-theory
variables is not amenable. For these two reasons we here take a
conservative approach, by restricting ourselves to solutions of
the equations (\ref{eomsig}) for {\em finite-dimensional} subspaces
of $\KE$ only, for which the `dictionary' between the $E_{10}$ and
the supergravity variables is known. We will keep the discussion as
simple as possible by only exciting one positive root, although, of course,
the analysis can be straightforwardly extended to more general
(but still finite-dimensional) subspaces.

To study the geodesic motion on finite-dimensional subspaces of the $\KE$
coset model, we consider a generator $E_\a$ of $E_{10}$ in the root space
of a positive root $\a=\sum_{i=1}^{10}m_i\a_i$ of the algebra. Our
labelling of the simple roots of $\E$ can be read off from
fig.~\ref{e10dynk}. Let $E_{-\a}=-\o(E_\a)$ where $\o$ is the Chevalley
involution \cite{Ka90}. Then we have the following relations
\be
\left[E_\a,E_{-\a}\right] = \sum_{i=1}^{10}m_i h_i \equiv h ,\quad\quad
\left[h, E_{\pm\a}\right] = \pm \a^2 E_{\pm\a}.
\ee
where $h$ lies in the Cartan subalgebra (CSA) of $E_{10}$. The generators
are normalised as follows
\be
\langle E_\a | E_{-\a} \rangle = 1,\quad\quad \langle h| h\rangle =
\a^2.
\ee
Since $\lae$ is a hyperbolic Kac--Moody algebra with symmetric generalized
Cartan matrix of signature $(-+\cdots +)$, $\a^2$ can take the values
$\a^2=2,0,-2,\ldots$. For $\a^2=2$ we have a standard $\mathfrak{sl}(2)$
algebra (in split real form). For $\a^2\le 0$ the simple root is
imaginary, and consequently we are dealing with a rank-one {\em Borcherds
algebra} with generalized Cartan matrix $(\a^2)$. Details on Borcherds
algebras can be found in \cite{Bo88} or \S11.13 of \cite{Ka90}.
For imaginary roots $\a$, the $\E$ root multiplicities
${\textrm{mult}}_{E_{10}}(\a)$ increase exponentially with $-\alpha^2$, and
we will thus have a large set of Lie algebra elements
$\{ E_{\a,s}|s=1,\dots, {\textrm{mult}}_{E_{10}}(\a)\}$ to choose from.
Below, however, we will need only one particular element from this
root space, and will therefore omit the label $s$.

We will thus consider a special two-dimensional subspace of the
(infinite-dimensional) coset $\KE$ coset, which is parametrized
as follows in triangular gauge\footnote{This is a standard
parametrization of the $SL(2)/SO(2)$ coset using the Iwasawa
decomposition. The same type of parametrization was also used for
BPS solutions in an $E_{11}$ context in
\cite{Englert:2003py} and in \cite{West:2004st,Cook:2004er}.}
\be
\cV(t) =
  e^{A(t)\cdot E_\a}\,e^{\phi(t)\cdot h}
  \quad\Longrightarrow\quad
  \cV^{-1}\pt\cV=\underbrace{e^{-\a^2\phi}\pt A}_{P_\a}\cdot E_\a
  + \pt\phi\cdot h.
\ee
The Lagrangian (\ref{sigmod}) evaluated on these fields is
\be
\cL = \frac{\a^2}2 (\pt\phi)^2 + \frac12 P_\a^2,
\ee
where we have adopted the affine gauge $n=1$. The equations of
motion (\ref{eomsig}) now read
\be\label{eom}
e^{-2\a^2\phi}\pt A &=& a\quad\quad\LRA\quad\quad P_\a = a e^{\a^2\phi},\nn\\
\pt^2\phi &=& -\frac12 a^2 e^{2\a^2\phi}.
\ee
The $A$ equation has already been integrated once using the
cyclicity of $A$, with $a$ the real constant of integration.
If $\a^2\ne 0$ we define
\be\label{2phis}
\vp(t) := \a^2\, \phi(t)
\ee
which obeys the Liouville equation
\be\label{eom2}
\pt^2\vp=- C e^{2\vp},\quad\quad C=\frac12 a^2\a^2.
\ee
Obviously, the sign of $C$ is positive if $\a$ is real and negative if
$\a$ is (pure) imaginary, {\em i.e.} time-like. Eq.~(\ref{eom2}) can
be integrated once, with one integration constant $E$
\be\label{eom3}
(\pt\vp)^2 + C e^{2\vp} = E \;\; ,
\ee
which is the energy of the effective one-dimensional system obtained
by integrating out $A(t)$. For $C\ne 0$,
Eq.~(\ref{eom3}) can be integrated explicitly
with the result (see e.g. \cite{DeIvMe05})
\be\label{vpfuns}
\vp(t) &=& - \ln\left[ \sqrt{\frac{C}{E}} \cosh(\sqrt{E}t)\right]\;\;\;
           \mbox{ for $C>0\, ,\, E>0$} \;\; ,\nn\\
\vp(t)&=& - \ln\left[ \sqrt{- \frac{C}{E}} \sinh(\sqrt{E}t)\right]\;\;\;
          \mbox{for $C<0\,,\, E>0$}\;\; ,\nn\\
\vp(t)&=& - \ln\left[\sqrt{-C}t\right] \qquad\qquad\quad\;\;
          \mbox{for $C<0\,, \,E=0$}\;\; , \nn\\
\vp(t) &=& - \ln\left[ \sqrt{\frac{C}{E}}\cos(\sqrt{-E}t)\right]\;\;\;
          \mbox{for $C<0\,,\, E<0$}.
\ee
(for vanishing time shifts). The time dependence of $A$ is then
easily deduced by integrating the equation $\partial_t A = a e^{2\vp}$.
The trajectories $\big( \vp(t), A(t)\big)$ are geodesics on the
non-compact coset spaces $SL(2)/SO(2)$ (for $C>0\LRA \a^2>0$) and
$SL(2)/SO(1,1)$ (for $C<0\LRA\a^2<0$).

If $C=0$, we must distinguish
two cases depending on the value $\a^2$. For $\a^2\ne 0$ (hence $a=0$),
the solution is a straight motion $\vp(t) = \pm\sqrt{E} t $, which
can be transformed back into a straight motion in the original $\phi$
space by eq.~(\ref{2phis}).\footnote{This is just the Kasner
  solution.}
If $\a^2=0$ (imaginary lightlike simple root), $\vp$ is not a valid
coordinate on the coset space and the solution for $\phi$ has to be computed
from eq.~(\ref{eom}), yielding $\phi(t)=-\frac12 a^2 t^2 + \sqrt{E}t$.

The above equations solve the $\s$-model equations of motion, but in
addition we have to satisfy the Hamiltonian constraint $\cH=\langle
\cP|\cP\rangle=0$ of the $\s$-model, according to eq.~(\ref{eomsig}).
We can compute the contribution
$\cH_\a$ of the above sector to $\cH$ for $\a^2\ne 0$ easily from (\ref{eom}):
\be\label{hamcontr}
\cH_\a = \frac12 a^2 e^{2\a^2\phi} + \a^2 (\pt\phi)^2 \equiv
   \frac12 a^2 e^{2\vp} + \frac1{\a^2}(\pt\vp)^2 = \frac{E}{\a^2}.
\ee
Evidently, unless $E=0$, more fields are required to satisfy the full
Hamiltonian constraint of the $\KE$ $\s$-model in order for the
resulting trajectory to be a {\em null} geodesic on the $\KE$ manifold.
There are many ways to achieve this; in all cases, we need at least one
timelike imaginary root (or at least the associated Cartan element) for
this. For the present purposes it is, however,
sufficient to superimpose two commuting
rank-one algebras with different signs for $E$. Commuting subalgebras
are convenient because the equations of motion decouple\footnote{In the
general case one would have to deal with a Toda-type system (if the
relevant algebra is still finite-dimensional).} with the following
result for the Hamiltonian
\be\label{sumham}
\cH = \sum_i \cH_{\a_i},
\ee
if $\a_i$ labels the different commuting subalgebras. A necessary
condition for two rank-one subalgebras to commute is $\langle \a_i |
\a_j \rangle =0$. For real roots $\a_i$ and $\a_j$ this is also a
sufficient condition.

\end{subsection}

\begin{subsection}{Example: Accelerated Cosmology}
\label{accsec}

We now illustrate this by exhibiting a solution with two commuting
subalgebras, one of which is $\mathfrak{sl}(2)$ and the other one is
Borcherds (although we only excite the $\mathfrak{u}(1)$ CSA part of it).
We will put tildes on all quantities belonging to the Borcherds
subalgebra. Concretely we take the two roots (written in a `root basis')
\be\label{apar}
\a&=& [ 0,0,0,0,0,0,0,0,0,1] \equiv \a_{10}  \nn\\
\ti{\a}&=& [6,12,18,24,30,36,42,28,14,21] \nn
\ee
so that
\be
\a^2=\langle \a|\a\rangle =2,\quad\quad
\ti{\a}^2= \langle \ti{\a}|\ti{\a}\rangle =-42\;\quad\quad \langle
\a|\ti{\a}\rangle=0\;.
\ee
The element $\ti{\a}$ is the fundamental weight \cite{Ka90} $\Lambda_7$
corresponding to node~7, and its multiplicity is ${\textrm{mult}}_{\E}
(\ti{\a})=4\,348\,985\,101$; its $A_9$ level is equal to the last
entry, whence $\ell = 21$. Correspondingly, the level of $\a$ is
$\ell=1$ and we identify the generator $E_\a$ with the component
$E^{8\,9\,10}$ of the rank three antisymmetric tensor
representation occurring at $\ell=1$ \cite{DaHeNi02,DaNi04}.
The commutator $[E_\a, E_{\ti{\a}}]$ will
belong to the root space of $\a+\ti{\a}$, whose multiplicity
is ${\textrm{mult}}_{\E}(\a+\ti{\a})=2\,221\,026\,189$.\footnote{For a
  list of $E_{10}$ multiplicities see e.g. \cite{Kl04}.} By
choosing $E_{\ti{\a}}$ appropriately we can therefore arrange that
this commutator
vanishes, so that the two algebras commute. The explicit expressions
for the Cartan generators corresponding to (\ref{apar})
in the $\mathfrak{gl}_{10}$ basis are
(cf. eqs.~(2.14) and (2.15) of \cite{DaNi04})
\be\label{hpar}
h &=& -\frac13 ({K^1}_1 + \dots + {K^7}_7) +
    \frac23 ({K^8}_8 + {K^9}_9 + {K^{10}}_{10}) \equiv h_{10} \nn\\
\ti{h} &=& {K^1}_1 + \dots + {K^7}_7
\ee

The identification of the $\E$ `matrix' $\cV$ with the supergravity
degrees of freedom is straightforward only for the diagonal components
of the metric (at a fixed spatial point). Abstractly, the dictionary
identifies the diagonal spatial metric as \cite{DaHeNi02}
\be\label{giden}
g_{mn}(t) = e^{2\phi(t)\cdot h + 2 \ti{\phi}(t)\cdot\ti{h}}\;.
\ee
In components, for the fields generators $h$ and $\ti{h}$ from
eq.~(\ref{hpar}) above this leads to
\be\label{metrsp}
g_{\ti{1}\ti{1}} = \cdots = g_{\ti{7}\ti{7}} =
e^{2\ti{\phi}-2\phi/3}\quad , \quad\quad g_{\ti{8}\ti{8}} =
g_{\ti{9}\ti{9}} = g_{\ti{10}\ti{10}} = e^{4\phi/3},
\ee
where tildes indicate that the corresponding indices are curved. The
$(tt)$-component of the metric in the $\s$-model correspondence
(in $n=1$ gauge) is
given as the square of the lapse function $N=\det({e_m}^a)$
\cite{DaNi04}, yielding
\be\label{metrtt}
g_{tt} = -e^{14\ti{\phi} - 2 \phi/3}.
\ee

For the off-diagonal $GL(10)$
generators and $\E$ generators outside $GL(10)$, on the other
hand, the precise correspondence is only known for low levels. Instead
of presenting here all known correspondences
\cite{DaHeNi02,KlNi04a,DaNi04,KlNi04b,KlNi05}, we present only an
identification we will need for $A_9$ level $\ell=1$
\be\label{lev1iden}
P^{(1)}_{abc} = F_{t\, abc} \;,
\ee
where $F_{t\, abc}$ is the 4-form field strength of $D=11$ supergravity,
with curved time index $t$ and {\em flat} spatial indices $a,b,c$.
The quantity on the l.h.s. multiplies the level-one generators $E^{abc}$
in the expansion of the Cartan form (\ref{cartform}).

We first consider the case $a\ne 0$ and $\ti{a}=0$, hence $C>0$ and
$\ti{C}=0$. Therefore the two energies of the solutions satisfy
$E>0$ and $\ti{E}\ge 0$, with the full Hamiltonian of the $\E$
model now evaluated to be
\be\label{enrel}
\cH \equiv \cH_\a + \cH_{\tilde\a}
= \frac12 E-\frac1{42}\ti{E} = 0 \quad \Longrightarrow\quad
  \ti{E} = 21 E.
\ee

We summarise the translation of the exact $\E$ solution (\ref{vpfuns})
with non-trivial fields $\phi$, $P_{8\,9\,10}$ and $\ti\phi$ to an
exact supergravity solution as
\be\label{accsol}
ds^2 &=& -e^{14\ti{\phi}-2\phi/3} dt^2 + e^{4\phi/3}(dx^2+dy^2+dz^2)
       +e^{2\ti{\phi}-2\phi/3} d\Sigma_{7,0}^2,\nn\\
F_{txyz} &=& \frac{E}{a \cosh^2(\sqrt{E} t)}=a\,e^{4\phi},
\ee
with
\be
\label{phiacc}
\phi(t)&=& -\frac12\ln\left[
     \frac{a}{\sqrt{E}}\cosh(\sqrt{E} (t-t_0))\right],\nn\\
\ti{\phi}(t) &=&  \frac1{42}\sqrt{\ti{E}}(t-t_1).
\ee
Here we have used eq.~(\ref{lev1iden}). Note also that the flat spatial
indices $8,9,10$ on the 4-form field strength have been converted
to {\em curved} indices $x,y,z$ in (\ref{accsol}).
The spatial coordinates of a four-dimensional space-time are now
designated as $x\equiv x_8, y\equiv x_9$ and $z\equiv x_{10}$,
and the $d\Sigma_{7,k}^2$ represents the metric of constant
curvature $k$ on a (compact) seven-dimensional manifold; here, however,
we restrict attention to the flat metric with $k=0$. This solution is
identical to the cosmological solution of
\cite{Ohta:2003pu,Wo03}.\footnote{\label{sbranefn}The solution
 (\ref{accsol}) is also equivalent to the SM2-brane solution,
see \cite{Oh04} and references therein.}
The solution (\ref{accsol}) exhibits a period of accelerated expansion
for the four-dimensional universe with coordinates $(t,x,y,z)$.
This was shown in \cite{Ohta:2003pu,Wo03} by a careful
analysis of the relevant `frames'. More precisely, in order to ascertain
whether or not the solution exhibits accelerated expansion, one must first
bring the metric of four-dimensional space-time into Einstein
conformal frame \cite{ToWo03} ({\em i.e.} no dependence on scalar fields
in front of the four-dimensional Einstein-Hilbert term). To be sure, a fully
realistic solution of this type remains to be found, but let us emphasize
again the important fact that M theory does admit cosmological solutions
with interesting profiles for the time evolution of the cosmic scale factor
in a rather natural manner.

\end{subsection}

\begin{subsection}{Curved internal spaces?}
\label{nonflat}

There exist analogues of the solution (\ref{accsol})
for the cases when $d\Sigma_{7,k}^2$ is the metric on a
compact Einstein space ${\cal M}_{7,k}$
with constant positive or negative scalar
curvature $R=\pm 42 k^2$. These spherical resp. hyperbolic
internal spaces have been studied in detail
in the literature \cite{Wo03,Oh03,Oh04}. The solution is in
form identical to (\ref{accsol}); only the function $\ti{\phi}$
changes its functional form from (\ref{phiacc}) to
\be\label{curvspaces}
\ti{\phi}(t)=\left\{\begin{array}{cl}
  -\frac16\ln\big[\frac{42k}{\sqrt{\ti{E}}}\cosh\big(
  \frac{\sqrt{\ti{E}}}{7}(t-t_1)\big)\big]&{\rm
  spherical\,\, space}\,\, R=42k^2\\&\\
  -\frac16\ln\big[\frac{42k}{\sqrt{\ti{E}}}\sinh\big(
  \frac{\sqrt{\ti{E}}}{7}(t-t_1)\big)\big]&{\rm
  hyperbolic\,\, space}\,\, R=-42k^2
  \end{array}\right.
\ee

We note that the second solution is very similar to the $C<0, E>0$ case
of eq.~(\ref{vpfuns}). This suggests that an analogous solution for
the $\KE$ coset model can be obtained by `switching on' the positive
step operator $E_{\ti{\a}}$ belonging to the Borcherds subalgebra, such
that $\ti{a}\ne 0$, and such that the imaginary $\ell=21$ root $\ti{\a}$ would
mimic the effect of the hyperbolic internal space. Indeed, for $\ti{a}\ne 0$
we also arrive at an {\em exact} solution of the $\KE$ model where now the
field $\ti{A}$, associated with a particular positive step operator
$E_\ti{\a}$ in
the $\ti{\a}$ root space, is obtained from integrating $\pt\ti{A}=\ti{a}
e^{2\ti{\vp}}$ and hence no longer constant. The function $\ti{\vp}$
here is the $C<0, E>0$ solution in (\ref{vpfuns}) and is functionally
of the same form as in (\ref{curvspaces}) for the hyperbolic case.
However, the solutions obtained for supergravity and the coset model
obtained in this way are no longer the same, because the coefficients of
the logarithms are different in (\ref{curvspaces}) and (\ref{vpfuns}).
Namely, for the $\E$ solution it is $\frac1{42}$ (after dividing
(\ref{vpfuns}) by $\ti{\a}^2$), as compared to $-\frac16$ in
(\ref{curvspaces}). This discrepancy also manifests itself in the
comparison of the Hamiltonian constraint for the $\E$ model and
the Hamiltonian constraint following from the $D=11$ supergravity
equations of motion, expressed in terms of the fields prior to
redefinition (\ref{2phis})
\be\label{hamdif}
2(\pt\phi)^2 + \frac12 a^2  e^{4\phi} - 42 (\pt\ti{\phi})^2 +
\frac12 {\ti{a}}^2 e^{-84 \ti{\phi}} &=& 0 \quad\quad (\E)\nn\\
2(\pt\phi)^2 + \frac12 a^2  e^{4\phi} - 42 (\pt\ti{\phi})^2 +
42 k^2 e^{12 \ti{\phi}} &=& 0 \quad\quad (D=11)
\ee
We emphasize again that we are dealing with two distinct exact solutions
of the $\KE$ model and of $D=11$ supergravity, respectively. From the
discrepancy in (\ref{hamdif}) we are led to conclude that we need to
take into account more positive step generators than simply $E_{\ti{\a}}$,
if we want to incorporate the solution with hyperbolic internal space
into the $\KE$ model. The above mismatch is therefore not entirely unexpected
since the required dependence on the coordinates of the compact space has to
be rather complicated in order to allow for constant negative curvature.

However, the spherical solution in (\ref{curvspaces}) is more puzzling.
Looking at the set of solutions (\ref{vpfuns}) to the Liouville
equation, we see that the
hyperbolic cosine only appears for real roots. The split of
coordinates in (\ref{accsol}) however suggests taking the CSA
elements $h$ and $\ti{h}$ of eq.~(\ref{hpar}) with $h$ real and
$\ti{h}$ imaginary. This is related to the question raised in
\cite{DaHeNi03} concerning the sign of contributions to the
Hamiltonian constraint in supergravity and in $\E$. There are
cosmological solutions to gravity which contribute negatively to
the Hamiltonian constraint. The spherical solutions with positive
constant curvature above is an example of such a cosmology. In
$\E$, on the other hand, all generators except for one direction
in the CSA have a positive contribution to the Hamiltonian
constraint. We have not been able to remedy this mismatch.

Let us at this point make one remark concerning the limit $k\to 0$
of the solutions (\ref{curvspaces}). In \cite{Wo03} it was noted
that the limit $k\to 0$ is not continuous and does not give the
correct flat space solution (\ref{phiacc}). This can be remedied
by noting that in the construction of the solution
(\ref{curvspaces}) the time shift appears as an integration
constant which is related to the curvature scale by $\ln k$.
Therefore, contrary to previous appearances, the limit $k\to 0$
{\em can} smoothly reproduce the flat space solution, via
\be
\lim_{k\to 0} \big[k \sinh(t-\ln k)\big] =\frac12 e^t,
\ee
leading to the functional behaviour of the flat case solution.
\end{subsection}

\end{section}

\begin{section}{Conclusion}
\label{concsec}
We have seen that the $\E$ equations can be solved rather easily for
dimension two subspaces, and that the solution to the resulting
Liouville equation translates into a (known) interesting cosmological
solution to the bosonic $D=11$ theory. The generalisation to
subspaces of dimension greater than two is straightforward, since
the resulting equations will be of Toda-type and can be integrated
in closed form (see for example \cite{RaSa97}). We stress again that
this abstract solution of geodesic motion on a group coset can be
interpreted in many different ways as solutions to the maximal
supergravity theories or their reductions by employing different
correspondences to type IIA, IIB, pure type I and pure Einstein
gravity which were detailed in
\cite{KlNi04a,KlNi04b,KlNi05}. In this way one finds from
(\ref{vpfuns}) all extremal S-branes. Their intersections rules
are also implemented by an orthogonality of two dimension two
subspaces, similar to the results of
\cite{Ivashchuk:1997pk,DeKa02,EnHoWe03,EnHo04b}. We note that our
approach does not
require the introduction of additional phantom fields unlike, for instance,
Ref. \cite{DeIvMe05} since there is a negative norm field
contained in both the $\KE$ model and the reduction of $D=11$.
The known dictionaries also cover one light-like (isotropic)
root of $\lae$, belonging to a Heisenberg subalgebra of $\lae$.
One can easily construct a solution using this Heisenberg subcoset
of $\KE$. This solution is a purely gravitational solution.

On the basis of the present results, it appears that the $\KE$ model
disfavors static solutions, such as $AdS_4\times S^7$. The reason is
that negative contributions to the Hamiltonian constraint always come
from the Cartan subalgebra, and the latter will only contribute if the
diagonal metric degrees of freedom depend on time. The situation is
somewhat reminiscent of `Einstein's dilemma', namely the impossibility of
finding static solutions for Einstein's gravitational field equations,
that led him to introduce a cosmological constant. However, it appears
that, for the $\E$ $\sigma$-model, such a remedy is not at
hand.\footnote{One can add a {\em one-dimensional} cosmological
  constant $\L_1$ to the $\KE$ Lagragian (\ref{sigmod}), thereby
  rotating the null
  geodesics to space-like or time-like geodesics depending on the sign
  of $\L_1$. However, the effect of this term is not the same as that
  of a higher-dimensional cosmological constant in the corresponding
  supergravity theory (if allowed by supersymmetry).}

The action of the $\E$ Weyl group $\cW$ on
these solutions should contain interesting features.
The orbit of a real root (like $\a$ in the
example of section \ref{accsec}) under $\cW$ covers {\em all}
positive roots and therefore relates solutions of different type:
The SM$2$ and SM$5$ solution are mapped into one another but also
into pure gravitational waves or monopole solutions. This is
well-known from U-duality \cite{HuTo95,ObPi99} but for $\E$
the action of $\cW$ also relates these solutions to as yet poorly
understood scenarios involving higher level generators.\\

\end{section}

{\bf Acknowledgements}\\
\noindent The authors thank M.~Henneaux and M.~Wohlfarth for clarifying
correspondence. AK would like to thank the organisers of the Young
researchers workshop in Crete for the opportunity to present the
results of this paper. This work was supported in part by the EU network
grant  MRTN-CT-2004 -- 512194.

\end{document}